# Electric and Magnetic Tuning Between the Trivial and Topological Phases in InAs/GaSb Double Quantum Wells

Fanming Qu[1,*], Arjan J. A. Beukman[1,*], Stevan Nadj-Perge[1], Michael Wimmer[1], Binh-Minh Nguyen[2], Wei Yi[2], Jacob Thorp[2], Marko Sokolich[2], Andrey A. Kiselev[2], Michael J. Manfra[3], Charles M. Marcus[4], Leo P. Kouwenhoven[1,§]

[1] QuTech and Kavli Institute of Nanoscience, Delft University of Technology, 2600 GA Delft, Netherlands

[2] HRL Laboratories, 3011 Malibu Canyon Rd, Malibu, CA 90265, USA

[3] Department of Physics, Purdue University, West Lafayette, Indiana 47907, USA

[4] Center for Quantum Devices, Niels Bohr Institute, University of Copenhagen, Copenhagen, Denmark

[*] These authors contributed equally to this work.

[§] l.p.kouwenhoven@tudelft.nl

**Among the theoretically predicted two-dimensional topological insulators, InAs/GaSb double quantum wells (DQWs) have a unique double-layered structure with electron and hole gases separated in two layers, which enables tuning of the band alignment via electric and magnetic fields. However, the rich trivial-topological phase diagram has yet to be experimentally explored. We present an *in situ* and continuous tuning between the trivial and topological insulating phases in InAs/GaSb DQWs through electrical dual-gating. Furthermore, we show that an in-plane magnetic field shifts the electron and hole bands relatively to each other in momentum space, functioning as a powerful tool to discriminate between the topologically distinct states.**

Two-dimensional topological insulators (2DTIs), known also as quantum spin Hall insulators, are a novel class of materials characterized by an insulating bulk and gapless helical edges [1-4]. Double quantum wells (DQWs) of indium arsenide and gallium antimonide (InAs/GaSb) have a unique type-II broken gap band alignment and are especially interesting since the electron and hole gases which form a topological band structure are spatially separated [5-8]. For the appropriate layer thicknesses, the top of the hole band in GaSb lies above the bottom of the electron band in InAs, hence for small momentum (around $k$ = 0) the band structure is inverted. At the crossing point ($k_{cross}$) of the two bands, coupling of the electrons and holes opens up a bulk hybridization gap [9-16] with gapless helical edge modes [5]. The size of the gap is determined by both $k_{cross}$ and the overlap of the electron and hole wave functions [17]. Due to the spatial separation of the two gases, electric and magnetic fields can induce relative shifts of the bands in energy and momentum [10, 18, 19], respectively. By controlling such shifts, it is possible to *in situ* tune between the trivial and topological insulating phases, which is the key advantage of InAs/GaSb compared to the other known 2DTIs [5, 20-22].

Here, for the first time, we map out the full phase diagram of the InAs/GaSb DQWs by independent control of the Fermi level and the band alignment through electric dual-gating. In particular, we observe the phase transition between the trivial insulator (normal gap) and topological insulator (hybridization gap). Moreover, the evolution of the resistance for in-plane magnetic fields is different in the two distinct phases, consistent with the fact that one is trivial, and the other topological.

In InAs/GaSb DQWs, the band alignment can be controlled by top and back gate electrodes [5, 18] (see the structure shown in Fig. 1(a)). The two gates control the perpendicular electric field $E_z$, which shifts the electron and the hole band relatively to each other in energy by $\Delta E = eE_z\langle z\rangle$ ($\langle z\rangle$ is the average separation of the electron and hole gases), and the position of the Fermi level $E_F$. The resulting trivial-topological phase diagram is shown in Fig. 1(b). On the left side of the vertical white line that marks the electric field for which the two bands touch, the DQWs are in the trivial regime.

Independently from the electric field, Fermi level can be tuned either to cross the electron band (point A), be in the normal gap (point B) or cross the hole band (point C). The red and blue background colors denote the electron and hole densities, respectively. By increasing $E_z$ the two bands move towards each other and the size of the normal gap decreases until the two bands touch at a certain value of $E_z$ (marked by the white line). By increasing $E_z$ further, the two bands overlap (invert) and the system enters the topologically non-trivial regime. In such inverted regime, the electrons and holes have the same density at $k_{cross}$ and mixing between them opens up a hybridization gap (white region around point H at the right side of the white line). The green lines separate the region with a single type of carriers (electrons or holes) from the region where both types of carriers are present. Figure 1(c) shows the phase diagram as a function of the back gate ($V_{BG}$) and top gate ($V_{TG}$) voltages (assuming the same coupling strength from both gates; see Supplemental Material Section III). The yellow dashed lines indicate constant $E_z$, and thus a fixed band alignment. Along the black dashed line ($V_{TG}=-V_{BG}$), the total carrier density is zero (charge neutrality). The red and blue curves represent the constant density lines for electrons and holes, respectively. These curves bend when the Fermi level starts to cross both electron and hole bands (along the green lines) as the total density of states increases and screening sets in. The constant electron and hole density lines bend differently according to the effective masses of the two types of carriers and the asymmetric quantum well structures. Note that, the phase diagram shown in Fig. 1(c) follows the calculations by Liu *et al*. [5] qualitatively.

Our heterostructure was grown by molecular-beam epitaxy [23] (see Supplemental Material Fig. S5). A 100 nm buffer layer was first grown on a doped GaSb substrate, followed by a 50 nm AlSb bottom barrier. The DQWs consist of 5 nm GaSb and 12.5 nm InAs, followed by a 50 nm AlSb top barrier and a 3 nm GaSb cap layer. Importantly, the GaSb substrate is lattice matched with the subsequent layers, which eliminates the requirement of a thick buffer layer compared to commonly used GaAs substrate and therefore enables a strong coupling between the back gate and quantum wells. Furthermore, for such choice of substrate, strain and the amount of dislocations are reduced resulting in record values of carrier mobility for this type of DQWs [23] (see also Supplemental Material Fig. S6). The Hall bars of 100 µm by 20 µm used in our measurements are chemically wet etched (inset of Fig. 2(a)). Ohmic contacts are fabricated by etching to the InAs layer prior to evaporation of Ti (50 nm)/Au (300 nm) layers. A sputtered 70 nm thick $Si_3N_4$ gate dielectric layer is used to isolate the Ti/Au top gate from the heterostructure. Longitudinal and Hall resistances are measured using standard lockin techniques at 300 mK unless otherwise stated. Two nominally identical devices are studied in detail.

First we map out the phase diagram of the InAs/GaSb DQWs by measuring the longitudinal resistance ($R_{xx}$) as a function of $V_{TG}$ and $V_{BG}$ for device #1 (Fig. 2(a)). The phase diagram reveals two regions of high resistance, labeled as I and II. Line L (R) crosses region I (II) and the corresponding resistance trace is shown in Fig. 2(b) (Fig. 2(c)). Region I has a maximum resistance of 8 kΩ, while the resistance in region II reaches 90 kΩ. The two regions touch around $V_{BG}$=-0.2 V and $V_{TG}$=-1.2 V, where the resistance shows a lower value of 4 kΩ, indicating a closing of a gap at this point in gate space. From this point, two less pronounced resistance peaks extend out (highlighted by the green lines; see Supplemental Material Fig. S7), indicating the onset of the coexistence of electrons and holes, as explained below. Note that, the finite conductance in the gapped regions I and II may result from disorder potential fluctuations in the bulk. In addition, for the inverted regime, level broadening will result in a finite residual bulk conductivity even at $T$=0 K [17]. The contribution from helical edge modes to the in-gap conductivity is expected to be small as the length of the Hall bar is significantly larger than the coherence length of the helical edge mode [13, 14].

To investigate the nature of the two gapped regions, we perform magnetoresistance measurements in perpendicular magnetic field ($B_z$) at the indicated points along lines L and R shown in Fig. 2(a). For clarity, five out of the seventeen measured Hall traces are shown in Figs. 3(a) and (b), for lines L and R, respectively. In Fig. 3(a), Hall traces at positions 1 (black) and 4 (red) taken along line L, start with a negative slope for small magnetic fields but bend up when magnetic field increases, indicating the coexistence of a majority of holes and a minority of electrons. Across the gap (e.g. point 12), the Hall trace also has a bend but a negative slope prevails, implying a majority of electrons. This interpretation can be clearly recognized in the extracted carrier types and densities along line L presented in Fig. 3(c).

In Fig. 3(c), the electron density (black squares) is obtained from the Shubnikov-de Haas (SdH) quantum oscillations which are present at all points along line L. These SdH oscillations are exclusively generated by the electrons in the InAs layer as a direct result of the much higher mobility for electrons (see Supplemental Material Fig. S6) [24]. To extract the hole densities, two different approaches are taken. First, at the left side, the hole concentration (blue solid triangles) is derived from a fit to the Hall traces using a two-carrier model. Such a fit also provides the electron density, as indicated by the red solid triangles (see Supplemental Material Section I). Second, on the right side of the gap, the hole density (blue open triangles) equals to the difference between the total density and the electron density (obtained from SdH oscillations). The total density is calculated from the Hall slope at high magnetic fields, since for $B \gg 1/\mu_e, 1/\mu_h$, $R_{xy} \approx \frac{B}{(p-n)e}$, where $\mu_e$ ($\mu_h$) is the electron (hole) mobility and $p$, $n$ are the hole density and the electron density, respectively. This analysis maps out both the electron and hole densities across gapped region I. The extrapolated hole density (brown dashed line) crosses the electron density near the center of the resistance peak at $n=p\approx 4\times 10^{15}$ m$^{-2}$. Accordingly, the wave vector at the crossing of the electron and hole bands is $k_{cross}=1.59\times 10^8$ m$^{-1}$ [13, 14]. The above analysis for gapped region I is consistent with an inverted band alignment and a hybridization gap opening at the crossing of the two bands.

We now turn to region II which is crossed by line R (Fig. 2(a)). In contrast to region I, the Hall traces along line R are nearly linear, as shown in Fig. 3(b). First, at the right side of the resistance peak, density values are obtained from the Hall slope (red dots) and the SdH period in $1/B$ (black squares). The close agreement between the Hall and SdH densities implies that solely electrons are present. Secondly, close to the gap, as the resistance becomes large and the carrier density is low, the Hall traces are characterized by fluctuations (such as at position 9). Finally, at the left side of the resistance peak, the transport is dominated by holes as obvious from the positive slope of the Hall traces. Importantly, in contrast to gapped region I, here both electron and hole densities are vanishingly small at the resistance peak, typical for a trivial insulator with a normal band gap. We note however that in the hole regime the Hall traces do have a slight bend. A two-carrier model is used to extract the hole density (blue solid triangles in Fig. 3(d)) and a small residual electron density of $\sim 5\times10^{14}$ m$^{-2}$, which may indicate a parallel conducting path.

To substantiate the above identification of the two gaps, we investigate the band alignment and Fermi level position for the distinct regions in gate space. We apply a 2 T perpendicular magnetic field and measure the phase diagram, as shown in Fig. 4. At the right side of the 2D map, the parallel lines correspond to SdH oscillations. The uniform spacing indicates a linear change of electron density as a function of both $V_{BG}$ and $V_{TG}$. The red curve follows a fixed Landau level, along which the electron density is constant. However, following this line towards the left, at position D the curve bends, indicating a coexistence of electrons and holes [24]. Such a bend arises when the Fermi level crosses the top of the hole band (green solid line). A similar effect happens when electrons come in at the hole side (green dashed line). The two green lines originating from point B follow the kinks on the constant density lines and separate the regions of single type and two types of carriers. The position of the green lines here is consistent with the two less pronounced resistance peaks in Fig. 2(a). Note that, at the gate regime just below gapped region I (between the green dashed line and the gap in Fig. 4), the observed SdH oscillations are primarily from electrons because of the lower mobility for holes than electrons, even though the holes have a higher density (see Supplemental Material Figs. S6 and S8). Importantly the phase diagram taken in a finite magnetic field (Fig. 4) as well as the one taken at zero magnetic field (Fig. 2(a)) are fully consistent with the interpretation that: (i) Along the resistance peak from A to B, the electron and hole bands approach each other in energy with the Fermi level lying in the middle of the normal gap; (ii) At point B the two bands touch and the transition from normal to inverted band alignment takes place; and (iii) Towards position C, the overlap increases and the Fermi level lies in the hybridization gap.

A further confirmation for the origin of the high resistance regions is the dependence of the resistance peaks on in-plane magnetic field [10, 19]. An in-plane magnetic field shifts the electron and hole bands in momentum (in the direction perpendicular to the magnetic field) by a relative amount of $\Delta k_y = eB_x \langle z \rangle / \hbar$, which is expected to reduce the hybridization gap but not the normal gap [10, 19]. To investigate this prediction, we focus on device #2 which was mounted in the plane of the two main axes (x and y) of a vector magnet. Device #2 is identical to device #1, in the sense of material, dimensions and all fabrication processes. The phase diagram for device #2 is nominally the same as device #1 (as shown in Supplemental Material Fig. S7). Figs. 5(a) and 5(b) show the in-plane magnetic field, $B_y$, dependence of the resistance for device #2 along the same lines L and R as device #1, respectively, while Figs. 5(c) and 5(d) show the $B_x$ dependence accordingly.

For the case of region I (line L), the height of the resistance peak decreases [25] and the peak position shifts slightly in gate space for both $B_y$ and $B_x$ (Fig. 5). The decrease of the resistance peak is anisotropic as it decays faster in $B_y$ (perpendicular to the transport direction). These observations are consistent with a relative shift of the two bands in momentum for the inverted regime. The anisotropy is expected since $B_x$ (parallel to transport direction) shifts the bands in $k_y$, and $B_y$ shifts the bands in $k_x$ [10, 19]. The inset of Fig. 5(a) ((c)) shows a sketch of the band structure along $k_x$ for finite $B_y$ ($B_x$) and the hybridization gap closes more quickly for $B_y$ because of the 'tilted' gap. The relative shift at 4 T is estimated to be $\Delta k = eB\langle z\rangle/\hbar$ =5.3×10$^7$ m$^{-1}$, which is smaller than $k_{cross}$. In clear contrast, for region II (line R) the resistance remains unchanged up to 9 T, although the same amount of band shift as region I is expected. This insensitivity on $B_x$ and $B_y$ demonstrates the trivial nature of gapped region II. Hence, the in-plane magnetic field dependence further corroborates the different types of the two gapped regions.

Finally, to give a rough estimate on the size of the hybridization gap, we take gate positions 6 and 10 in Fig. 2(a) as the gap edges. Combining the electron density difference between the two selected points based on a linear extrapolation of $n_{SdH}$ and the constant 2D density of states $m_e/\pi\hbar^2$, we estimate a gap size of $\Delta= \pi\hbar^2(n_{10} - n_6)/m_e$=9.3 meV, which is larger than values reported in the literature [6, 10, 13-16]. The effective mass of $m_e$=0.04$m_0$ for electrons is deduced from the temperature dependence of the amplitude of SdH oscillations, i.e., the Dingle plot [26], deep in the electron regime ($V_{BG}$=1 V, $V_{TG}$=2 V) where $n$=2.2×10$^{16}$ m$^{-2}$ and $\mu_e$=70 m$^2$/Vs (see Supplemental Material Section II). The relatively large deduced gap may be overestimated due to the inaccuracy of the selected gap edge positions, or it is indeed large because of strong coupling between electrons and holes for these gate values. We also notice that at the upper-left side of gapped region I (larger $k_{cross}$) in Fig. 2(a) the peak resistance drops, indicating a decrease of the hybridization gap. Such decrease is presumably due to a reduced wave function overlap and hence a reduced coupling strength, in spite of the increased $k_{cross}$. From the Dingle plot deep in the electron regime we extract a quantum scattering time of $\tau_q$=0.27 ps, corresponding to a quantum level broadening [13, 14] of $\Gamma_e = \hbar/2\tau_q$=1.22 meV. However, close to the gap, the electron mobility drops by more than one order of amplitude ($\mu_e$=5.5 m$^2$/Vs at position 5), suggesting a much larger $\Gamma_e$. Hence, the total level broadening will be $\Gamma = \Gamma_e + \Gamma_h \gg$ 1.22 meV (where $\Gamma_h$ represents the hole contribution), which could account for the relatively low resistance at the hybridization gap.

In conclusion, we explored the full phase diagram of InAs/GaSb DQWs structure by measuring dual-gated Hall bars. We observed two gapped regimes manifested by regions with a longitudinal resistance peak. For one gapped region, the extracted electron and hole densities both vanish at the resistance peak. While for the other gapped region finite equal electron and hole densities are present around the resistance peak. Our findings are fully consistent with a scenario that one gapped region corresponds to a trivial insulator with a normal gap while the other corresponds to a topological insulator with a hybridization gap. Moreover, the dependence of the two resistance peaks on in-plane magnetic fields further corroborates the different origins of the two gapped regions.

**Acknowledgements**

We gratefully acknowledge Folkert de Vries, Rafal Skolasinski, Fabrizio Nichele, Morten Kjærgaard and Henri Suominen for very helpful discussions. This work has been supported by funding from the Netherlands Foundation for Fundamental Research on Matter (FOM), Microsoft Corporation Station Q and Danish National Research Foundation. S. N-P. acknowledges support of the European Community through the Marie Curie Fellowship (IOF 302937).

**References**

[1] C. Kane, and E. Mele, Phys. Rev. Lett. **95**, 226801 (2005).

[2] B. Bernevig, and S.-C. Zhang, Phys. Rev. Lett. **96**, 106802 (2006).

[3] B. A. Bernevig, T. L. Hughes, and S. C. Zhang, Science **314**, 1757 (2006).

[4] M. Konig *et al.*, Science **318**, 766 (2007).

[5] C. X. Liu *et al.*, Phys. Rev. Lett. **100**, 236601 (2008).

[6] I. Knez, R. R. Du, and G. Sullivan, Phys. Rev. Lett. **107**, 136603 (2011).

[7] I. Knez *et al.*, Phys. Rev. Lett. **112**, 026602 (2014).

[8] L. Du *et al.*, arXiv:1306.1925 (2013).

[9] S. de-Leon, L. Shvartsman, and B. Laikhtman, Phys. Rev. B **60**, 1861 (1999).

[10] M. Yang *et al.*, Phys. Rev. Lett. **78**, 4613 (1997).

[11] M. Altarelli, Phys. Rev. B **28**, 842 (1983).

[12] M. Lakrimi *et al.*, Phys. Rev. Lett. **79**, 3034 (1997).

[13] I. Knez, R. Du, and G. Sullivan, Phys. Rev. B **81**, 201301 (2010).

[14] F. Nichele *et al.*, Phys. Rev. Lett. **112**, 036802 (2014).

[15] K. Suzuki *et al.*, Phys. Rev. B **87**, 235311 (2013).

[16] W. Pan *et al.*, Appl. Phys. Lett. **102**, 033504 (2013).

[17] Y. Naveh, and B. Laikhtman, EPL **55**, 545 (2001).

[18] Y. Naveh, and B. Laikhtman, Appl. Phys. Lett. **66**, 1980 (1995).

[19] K. Choi *et al.*, Phys. Rev. B **38**, 12362 (1988).

[20] W. Yang, K. Chang, and S.-C. Zhang, Phys. Rev. Lett. **100**, 056602 (2008).

[21] I. Knez, R.-R. Du, and G. Sullivan, Phys. Rev. Lett. **109**, 186603 (2012).

[22] C.-C. Liu, W. Feng, and Y. Yao, Phys. Rev. Lett. **107**, 076802 (2011).

[23] B.-M. Nguyen *et al.*, Appl. Phys. Lett. **106**, 032107 (2015).

[24] L. Cooper *et al.*, Phys. Rev. B **57**, 11915 (1998).

[25] By measuring the Hall resistance, a ~7 degree angle between $B_y$ and the surface of the device is determined. As perpendicular magnetic field induces a large positive magnetoresistance, the peak resistance increases when $B_y$ is small as a result of the competion between the positive magnetoresistance and the gap closing, as shown in Fig. 5(a).

[26] T. Ihn, *Semiconductor nanostructures* (Oxford University Press, New York, 2010).

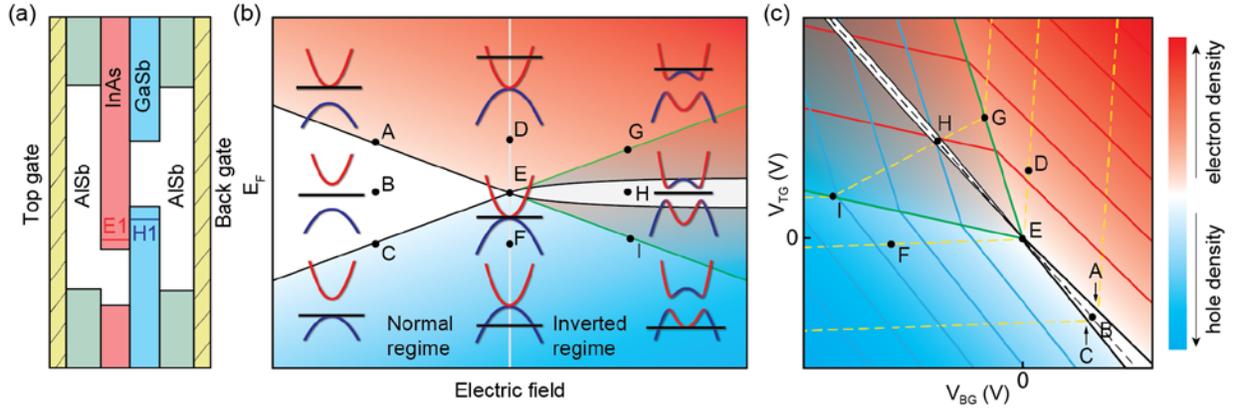

FIG 1: (color online) (a) Schematic of the InAs/GaSb DQWs structure. E1 and H1 mark the bottom of the conduction band and the top of the valence band respectively, showing an inverted band alignment. (b) Phase diagram as a function of the applied electric field and the Fermi level position ($E_F$), which can both be tuned by dual-gating using top gate and back gate. The insets show the band structures and the position of Fermi level at different points, A to I. The red and blue background colors represent the electron and hole densities, respectively. The vertical white line, along which the electron (red) and hole (blue) bands touch, separates the normal and inverted band alignment regimes. (c) Sketch of the phase diagram as a function of back gate voltage ($V_{BG}$) and top gate voltage ($V_{TG}$) (see Supplemental Material Section III). The labelled points in (b) are indicated accordingly in (c). The red and blue lines mark constant electron and hole densities, respectively. The yellow dashed lines indicate constant band overlap for the inverted case or constant band separation for the normal case. The black dashed line represents charge neutrality.

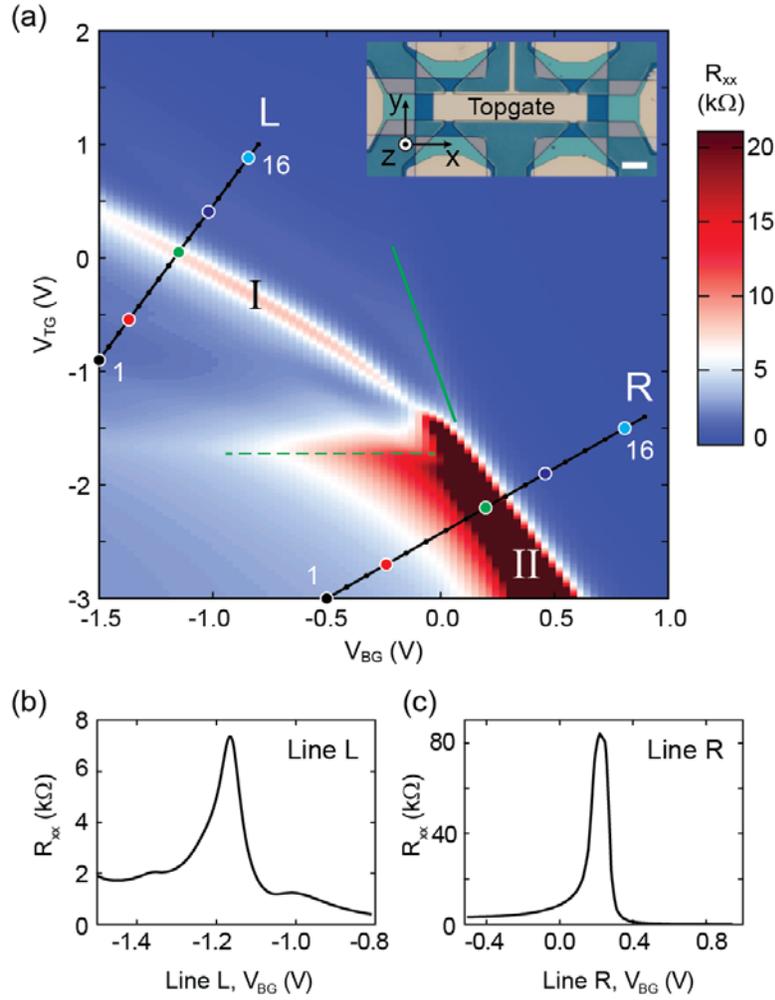

FIG 2: (color online) (a) Four terminal longitudinal resistance ($R_{xx}$) as a function of $V_{BG}$ and $V_{TG}$ for device #1 measured at 300 mK, showing the phase diagram of the InAs/GaSb DQWs. Lines L and R cross the two different gapped regions (resistance peaks), labeled as I and II. Colored dots indicate the positions in gate-space where longitudinal resistance and Hall traces are taken, as shown in Figs. 3(a) and (b). The two green lines indicate the two less pronounced resistance peaks (see text). Inset shows the optical image of the Hall bar. The scale bar represents 20 μm. (b), (c) Resistance along lines L and R, respectively.

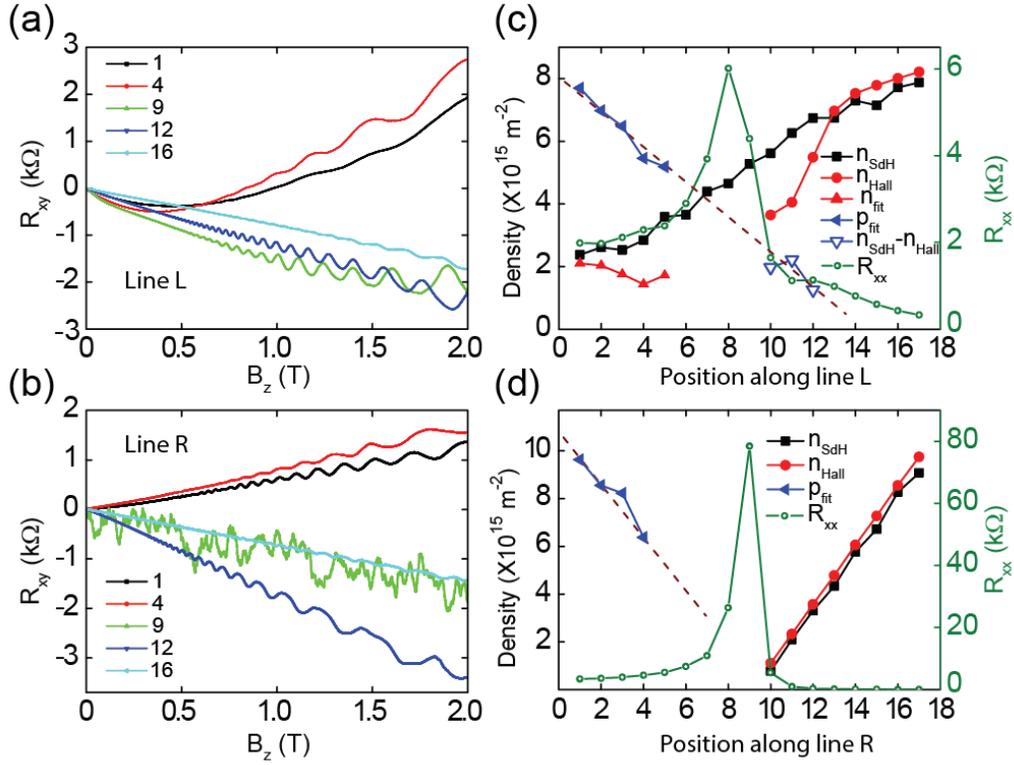

FIG 3: (color online) (a), (b) Hall resistance as a function of perpendicular magnetic field for five out of seventeen measured points along lines L and R in Fig. 2(a), respectively. The colors of the traces here correspond to the colored dots in Fig. 2(a). (c), (d) Carrier densities and longitudinal resistance for the seventeen uniformly distributed positions along lines L and R in Fig. 2(a), respectively. Black squares represent the electron densities ($n_{SdH}$) obtained from SdH. Red dots are the total densities ($n_{Hall}$) extracted from the Hall slope. Zero magnetic field resistances ($R_{xx}$) are shown by the green circles. In (c), the blue open triangles show the difference between the electron density and the total density ($n_{SdH}-n_{Hall}$), i.e., the hole density. At the left side of the resistance peak, the blue solid triangles and the red solid triangles represent the hole density ($p_{fit}$) and the electron density ($n_{fit}$) obtained from the fit of the bended Hall traces with a two-carrier model, respectively. In (d), the hole density near gapped region II deduced from fitting is shown by the blue solid triangles.

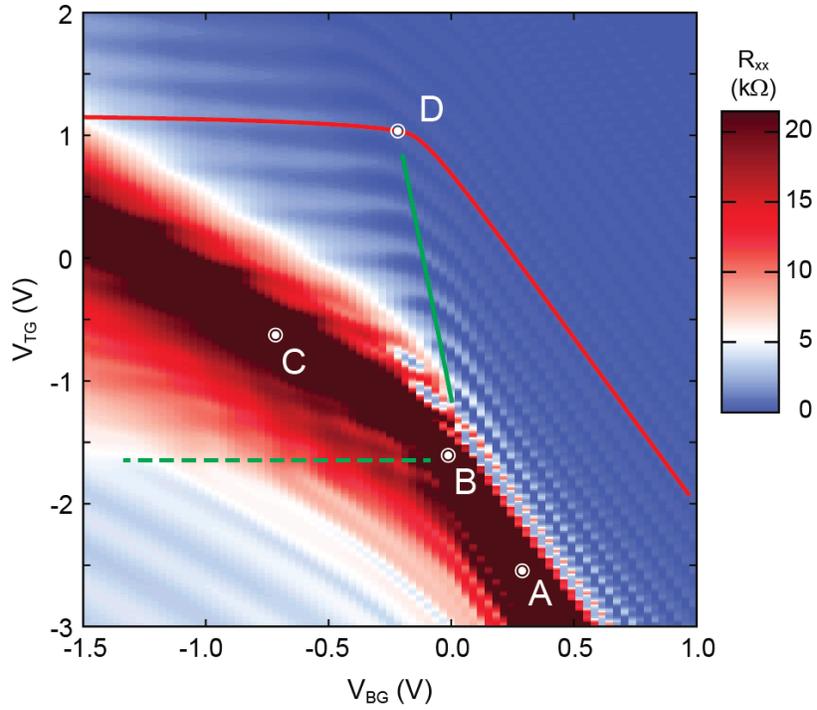

FIG 4: (color online) Longitudinal resistance as a function of $V_{BG}$ and $V_{TG}$ at a perpendicular magnetic field of 2 T for device #1. From A to B, the electron and hole bands get closer and an insulator to semimetal transition occurs. From B to C, in the inverted regime, the overlap of the two bands increases and the Fermi level lies in the hybridization gap. The red curve denotes a constant density line for electrons, where the Fermi level is fixed relative to the bottom of the electron band and only the hole band moves. The two green lines separate the regions with a single type and two types of carriers.

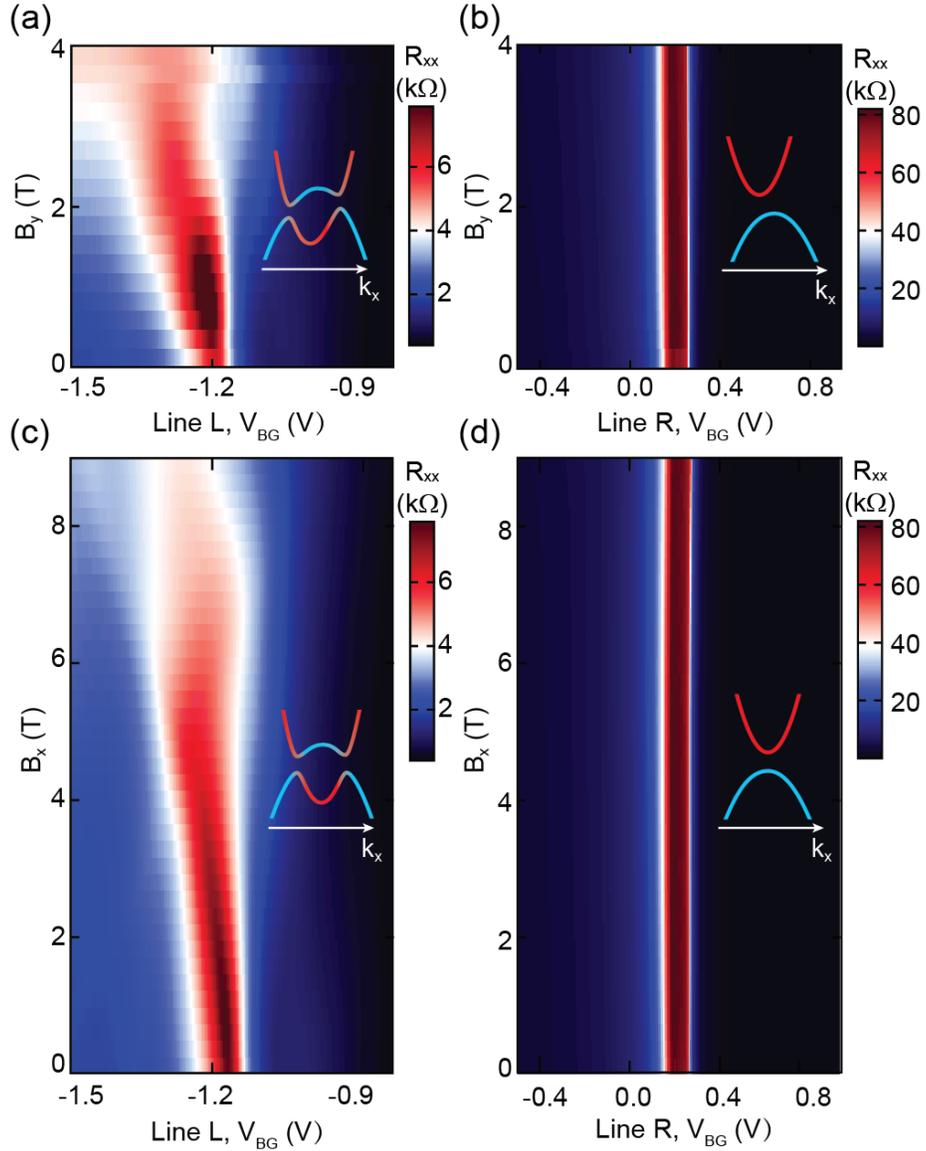

FIG 5: (color online) (a), (b) In-plane magnetic field ($B_y$) dependence of the longitudinal resistance for device #2, along the same lines L and R in Fig. 2(a). (c), (d) The same as (a), (b) but for $B_x$. The resistance for line L drops as $B_x$ or $B_y$ increases, while for line R, it stays constant. Insets in (a)-(d) show the band structure along $k_x$. As the relative shift of the bands in momentum is perpendicular to the in-plane magnetic field, the hybridization gap closes more quickly in $B_y$ (perpendicular to the transport direction) than in $B_x$ (parallel to the transport direction).

# Supplemental Material for 'Electric and Magnetic Tuning Between the Trivial and Topological Phases in InAs/GaSb Double Quantum Wells'

Fanming Qu[1,*], Arjan J. A. Beukman[1,*], Stevan Nadj-Perge[1], Michael Wimmer[1], Binh-Minh Nguyen[2], Wei Yi[2], Jacob Thorp[2], Marko Sokolich[2], Andrey A. Kiselev[2], Michael J. Manfra[3], Charles M. Marcus[4], Leo P. Kouwenhoven[1,§]

[1] QuTech and Kavli Institute of Nanoscience, Delft University of Technology, 2600 GA Delft, Netherlands
[2] HRL Laboratories, 3011 Malibu Canyon Rd, Malibu, CA 90265, USA
[3] Department of Physics, Purdue University, West Lafayette, Indiana 47907, USA
[4] Center for Quantum Devices, Niels Bohr Institute, University of Copenhagen, Copenhagen, Denmark
* These authors contributed equally to this work.
§ l.p.kouwenhoven@tudelft.nl

**I. Two-carrier model**

When electron and hole gases coexist in a two-dimensional system, the dependence of Hall resistance $R_{xy}$ on perpendicular magnetic field $B$ is [1]:

$$R_{xy} = \frac{-B[(n\mu_e^2 - p\mu_h^2) + B^2 \mu_e^2 \mu_h^2 (n-p)]}{e[(n\mu_e + p\mu_h)^2 + B^2 \mu_e^2 \mu_h^2 (n-p)^2]}$$

where $n$ and $p$ are the densities for electrons and holes, respectively; $\mu_e$ and $\mu_h$ are the corresponding mobilities. The measured Hall trace can be fitted using the formula above, and a constraint is given by the zero-field longitudinal resistance:

$$B = 0: \quad R_{xx} = \frac{L}{e(n\mu_e + p\mu_h)W}$$

where $L$ and $W$ are the length and width of the Hall bar, respectively.
In the limit of $B \gg 1/\mu_e, 1/\mu_h$, $R_{xy} \approx \frac{B}{(p-n)e}$.

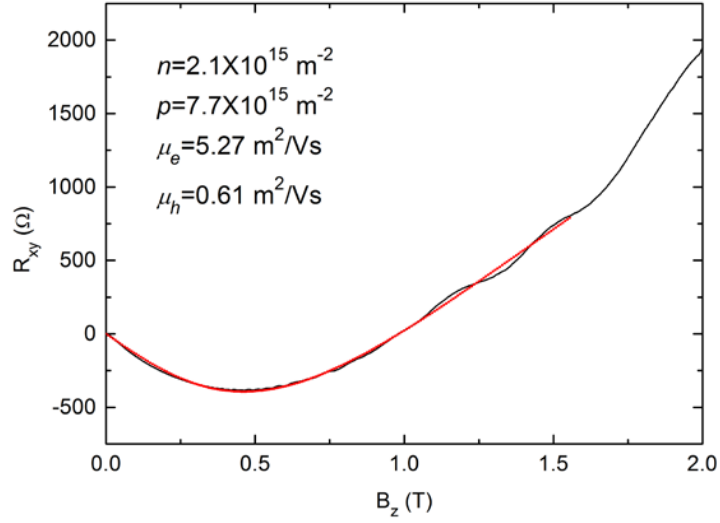

Figure S1: An example of two-carrier model fitting (red) of the Hall trace (black) taken at position 1 on line L in Fig. 2(a) in the main text. The zero-field longitudinal resistance is used as the constraint. The extracted carrier densities and mobilities are shown in the figure.

**II. Dingle plot**

For a two-dimensional electron gas in a perpendicular magnetic field $B$, the amplitude of the Shubnikov-de Haas (SdH) oscillations, $\Delta R$, is given by [2]:

$$\Delta R = 4R_0 \exp\left(\frac{-\pi m_e}{eB\tau_q}\right) \frac{2\pi^2 m_e k_B T}{e\hbar B} / \sinh\left(\frac{2\pi^2 m_e k_B T}{e\hbar B}\right)$$

where $R_0$ is the zero-field resistance, $m_e$ the effective electron mass, and $\tau_q$ the quantum scattering time. At a fixed magnetic field $B$, by measuring the temperature dependence of $\Delta R$, $m_e$ and $\tau_q$ can be extracted from the fit, the so called Dingle plot.

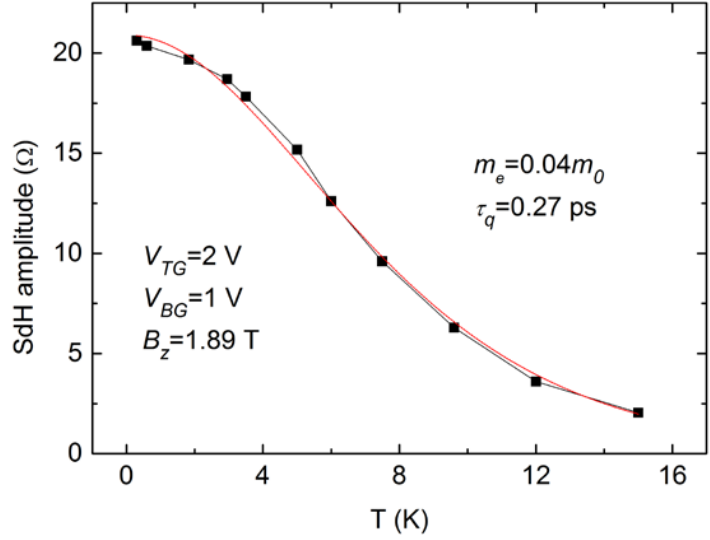

*Figure S2: Dingle plot taken deep in the electron regime, where $V_{TG}$=2 V and $V_{BG}$=1 V for device #1 in the main text. The fit (red) of the temperature dependence of the SdH oscillation amplitude (black) gives the effective electron mass $m_e$=0.04$m_0$ ($m_0$ is the electron mass), and the quantum scattering time $\tau_q$=0.27 ps. (Deep in the hole regime, the Dingle plot gives the effective hole mass $m_h$=0.09$m_0$.)*

**III. Calculation of the phase diagram**

A simple capacitor model (Fig. S3) is used to simulate the phase diagram of the InAs/GaSb double quantum wells (DQWs). The result (Fig. S4) is in qualitative agreement with the calculations by Liu et al. [3].

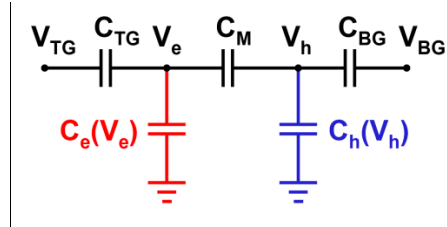

*Figure S3: Capacitor model of the InAs/GaSb double quantum wells device.*

The device operation can be explored and semi-quantitatively understood by considering electrostatics of its stack in the simplistic *equivalent capacitance model* [4] which is shown in Fig. S3. Here $C_{TG}$, $C_M$, and $C_{BG}$ are *geometric* capacitances between the top gate and the two-dimensional electron gas (2DEG) (or, more accurately, 2DEG electron density center plane located inside the InAs layer), between 2DEG and two-dimensional hole gas (2DHG), and between 2DHG and the back gate, respectively. $C_e$ and $C_h$ are *quantum* capacitances reflecting an energy penalty associated with the gradual filling of electron/hole states. It is proportional to the density of states $D$. For a 2D subband in a parabolic approximation, $D = |m|/\pi\hbar^2$ is a constant defined by the subband in-plane effective mass $m$. The penalty is zero when the corresponding subband is fully depleted, as $D=0$ in this case. With the Fermi surfaces of electron and hole 2D gases grounded through the Ohmic contacts, electron and hole accumulation can be parameterized by the energies of the electron and hole subband extrema, $-|e|V_e$ and $-|e|V_h$. Thus, $C_{e,h}$ are *piecewise constant* functions of $V_{e,h}$: $C_e=e^2 D_e$ when $V_e>0$ (and is zero otherwise), while $C_h=e^2 D_h$ only when $V_h<0$. This model can be numerically solved for $V_{e,h}$ at arbitrary back and top gate biases ($V_{BG}, V_{TG}$), to obtain the electron ($n=C_e V_e/|e|$) and hole ($p=-C_h V_h/|e|$) densities.

This simplistic model reproduces the anticipated physics well, resolving conditions in gate space when a normal gap forms (i.e., no conduction: $n=p=0$), single carrier situations (either $n=0$ or $p=0$), and when both electrons and holes are present simultaneously. The resulting phase diagram is shown in Fig. S4. It marks electron and hole depletion boundaries $V_{e,h}=0$ (green lines), charge neutrality line $n=p$ (dashed), and a zero-gap condition $V_e=V_h$ (yellow line). Constant density contours for both electrons and holes are also shown. Geometric capacitances $C_{TG}=3.6\times10^{11}|e|V^{-1}cm^{-2}$ =58 nF cm$^{-2}$, $C_M=9.8\times10^{12}|e|V^{-1}cm^{-2}$ =1.6 uF cm$^{-2}$, and $C_{BG}=4.9\times10^{11}|e|V^{-1}cm^{-2}$ =79 nFcm$^{-2}$ have been estimated for the actual multilayer dielectric/semiconductor stack using $\varepsilon_{SiN}=7.0$, $\varepsilon_{GaSb}=15.7$, $\varepsilon_{AlSb}=10.9$, and $\varepsilon_{InAs}=15.5$ dielectric constants, assuming that electron and hole central planes coincide with centers of InAs and GaSb layers, and that $V_{BG}$ is applied to the edge of the n-doped GaSb substrate. Experimentally determined $m_e=0.04m_0$ and $m_h=0.09m_0$ have been used to set quantum capacitance values at $C_e=1.7\times10^{13}|e|V^{-1}cm^{-2}$ =2.7 uFcm$^{-2}$ and $C_h=3.8\times10^{13}|e|V^{-1}cm^{-2}$ =6.0 uFcm$^{-2}$.

The above simple model does not include $C_{e,h}$ modulation due to electron and (especially) hole non-parabolicities, the spatial shift of the wave functions inside the InAs and GaSb layers, as well as other details related to the band alignment in the device.

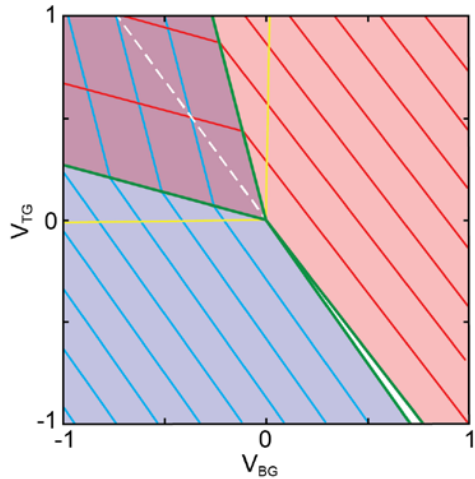

*Figure S4: Simulation result from the capacitor model. Red and blue shaded areas indicate regions of purely electrons and holes, respectively. The purple shaded area indicates a region where electrons and holes coexist. Red and blue lines represent equal density of electrons and holes, respectively. The yellow lines indicate points where the two bands touch each other. The white region at the bottom right part of the figure indicates the normal insulating gap. The dotted white line at the top left part connect points where electrons and holes have equal density and the hybridization gap is expected.*

## IV. Supplementary Figures

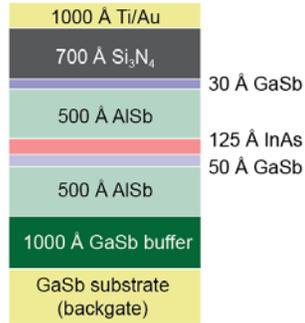

*Figure S5: Cross-section of the InAs/GaSb DQWs used in the main text. The quantum well contains a 12.5 nm InAs layer on top of a 5 nm GaSb layer. The structure is grown on a doped GaSb substrate.*

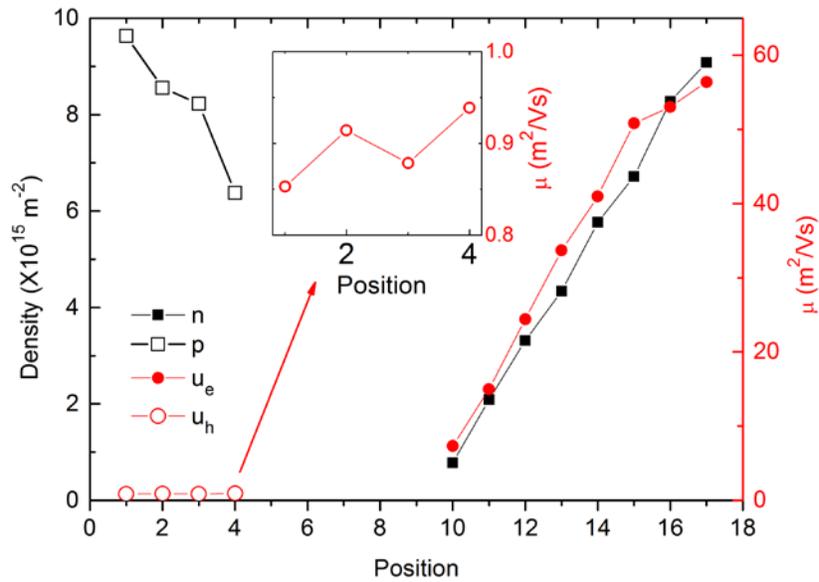

*Figure S6: Densities and mobilities for electrons and holes for device #1 along line R in Fig. 2(a) in the main text. The electron mobility is more than one order of magnitude higher than the hole mobility. Because of the application of the GaSb substrate, record values of the electron mobility are achieved.*

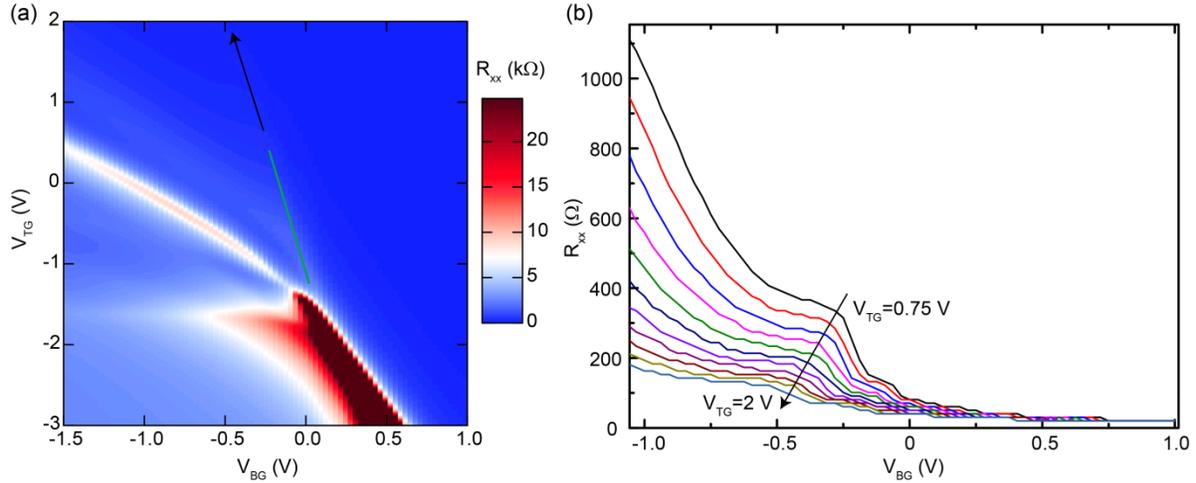

*Figure S7: (a) Longitudinal resistance as a function of back gate and top gate voltages for device #2 in the main text measured at 300 mK, which shows nominally the same phase diagram as Fig. 2(a) for device #1 in the main text. The green line indicates the onset of the coexistence of electrons and holes. (b) Line cuts taken from (a) between $V_{TG}$=0.75 V and 2 V. The arrow shows the shift of the small resistance peak, consistent with the arrow in (a).*

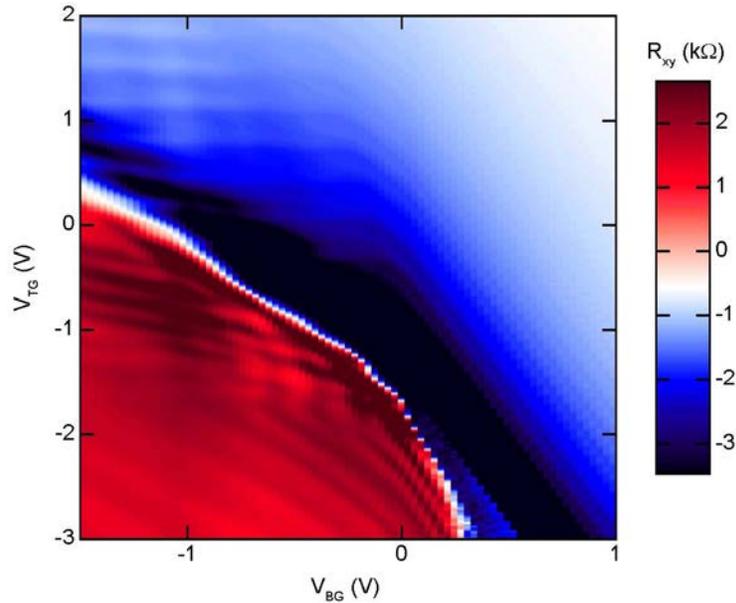

*Figure S8: Hall resistance as a function of both top gate and back gate voltages at a fixed perpendicular magnetic field of 2 T for device #1 in the main text. This figure was measured simultaneously with Fig. 4 in the main text. For regions with only holes or a majority of holes, the Hall resistance shows positive values, as shown by the red color. In contrast, the blue color indicates regions with only electrons or a majority of electrons.*

**References:**

[1] R. Smith, *Semiconductors* (Cambridge University Press, Cambridge, 1978).
[2] T. Ihn, *Semiconductor nanostructures* (Oxford University Press, New York, 2010).
[3] C. X. Liu *et al.*, Phys. Rev. Lett. **100**, 236601 (2008).
[4] S. Luryi, Appl. Phys. Lett. **52**, 501 (1988).
7